\documentclass[aps,prl,twocolumn,superscriptaddress,floatfix]{revtex4}
\usepackage{amsmath}
\usepackage{amsfonts}
\usepackage{amssymb}
\usepackage{graphicx}
\usepackage{cancel}
\usepackage[switch,columnwise]{lineno}
\usepackage{titlesec}
\usepackage{xcolor}
\usepackage{hyperref}
\usepackage[none]{hyphenat}

\makeatletter
\renewcommand{\section}{\@startsection{section}{1}{0mm}
  {-\baselineskip}{0.5\baselineskip}{\bf\leftline}}

\renewcommand{\subsection}{\@startsection{section}{1}{0mm}
  {-\baselineskip}{0.5\baselineskip}{\bf\leftline}}
\makeatother

\begin{document}

\title{Measurements with prediction and retrodiction on the collective spin of $10^{11}$ atoms beat the standard quantum limit
}

\author{Han Bao}%
\affiliation{Department of Physics, State Key Laboratory of Surface Physics and Key Laboratory of Micro
and Nano Photonic Structures (Ministry of Education), Fudan University, Shanghai 200433, China}%
\affiliation{Collaborative Innovation Center of Advanced Microstructures, Nanjing 210093, China}%
\author{Junlei Duan}%
\affiliation{Department of Physics, State Key Laboratory of Surface Physics and Key Laboratory of Micro
and Nano Photonic Structures (Ministry of Education), Fudan University, Shanghai 200433, China}%
\author{Xingda Lu}%
\affiliation{Department of Physics, State Key Laboratory of Surface Physics and Key Laboratory of Micro
and Nano Photonic Structures (Ministry of Education), Fudan University, Shanghai 200433, China}%
\author{Pengxiong Li}%
\affiliation{Department of Physics, State Key Laboratory of Surface Physics and Key Laboratory of Micro
and Nano Photonic Structures (Ministry of Education), Fudan University, Shanghai 200433, China}%
\author{Weizhi Qu}%
\affiliation{Department of Physics, State Key Laboratory of Surface Physics and Key Laboratory of Micro
and Nano Photonic Structures (Ministry of Education), Fudan University, Shanghai 200433, China}%
\author{Shenchao Jin}%
\affiliation{Department of Physics, State Key Laboratory of Surface Physics and Key Laboratory of Micro
and Nano Photonic Structures (Ministry of Education), Fudan University, Shanghai 200433, China}%
\author{Mingfeng Wang}%
\affiliation{Department of Physics, Wenzhou University, Zhejiang 325035, China}%
\author{Irina Novikova}%
\affiliation{Department of Physics, College of William and Mary, Williamsburg, Virginia 23187, USA}
\author{Eugeniy E. Mikhailov}%
\affiliation{Department of Physics, College of William and Mary, Williamsburg, Virginia 23187, USA}
\author{Kai-Feng Zhao}%
\affiliation{Applied Ion Beam Physics Laboratory, Key Laboratory of the Ministry of Education, and Institute of Modern Physics, Fudan University, Shanghai 200433, China}
\author{Klaus M\o lmer}%
\email{moelmer@phys.au.dk}
\affiliation{Department of Physics and Astronomy, Aahus University, Ny Munkegade 120, DK-8000 Aarhus C. Denmark}%
\author{Heng Shen}%
\email{heng.shen@physics.ox.ac.uk}
\affiliation{Clarendon Laboratory, University of Oxford, Parks Road, Oxford, OX1 3PU, UK}%
\author{Yanhong Xiao}%
\email{yxiao@fudan.edu.cn}
\affiliation{Department of Physics, State Key Laboratory of Surface Physics and Key Laboratory of Micro
and Nano Photonic Structures (Ministry of Education), Fudan University, Shanghai 200433, China}%
\affiliation{Collaborative Innovation Center of Advanced Microstructures, Nanjing 210093, China}
%
\begin{abstract}
Quantum probes using $N$ uncorrelated particles give a limit on the measurement sensitivity referred to as the standard quantum limit (SQL). The SQL, however,  can be overcome by exploiting quantum entangled states, such as spin squeezed states. We report generation of a quantum state, that surpasses the SQL for probing of the collective spin of $10^{11}$ $\text{Rb}$ atoms contained in a vapor cell. The state is prepared and verified by sequences of stroboscopic quantum non-demolition (QND) measurements, and we apply the theory of past quantum states to obtain the spin state information from the outcomes of both earlier and later QND measurements. In this way, we obtain a conditional noise reduction of 5.6 dB, and a metrologically-relevant squeezing of $4.5\pm0.40~\text{dB}$. The past quantum state yields tighter information on the spin component than we can obtain by a conventional QND measurement. Our squeezing results are obtained with 1000 times more atoms than in any previous experiments with a corresponding record $4.6\times10^{-13} rad^2$ variance of the angular fluctuations of a squeezed collective spin.
\end{abstract}
\maketitle

Measurements constitute the foundations of physical science. The aim of high-precision metrology is to reduce uncertainties and make the tightest possible conclusions from measurement data \cite{Lloyd}. Quantum systems are described by wave functions or density matrices, which yield probabilistic measurement outcomes, and for a continuously monitored system, the well-established theory of quantum trajectories employs stochastic master equations to describe the evolution with time of the density matrix $\rho(t)$, governed by the system Hamiltonian, dissipation, and effects 
associated with the measurements \cite{Wiseman-book}. For the case of Gaussian states and operations, the theory simplifies to equations for mean values and covariances, equivalent to classical Kalman filter theory \cite{Klaus2004}.

Knowing the value of $\rho(t)$, we may predict the outcome of the subsequent measurement on the system, and if QND probing has led to a state with reduced uncertainty of a specific observable, we may thus make an improved prediction of the subsequent measurement. Also, later measurements will have outcomes correlated with the present and previous ones, and the same way as daily life experience teaches us about past events and facts,
one may ask if it is possible in a quantum experiment to obtain more knowledge about a quantum state by using both earlier and later observations on a system. Such retrodiction was introduced first in the context of pre- and post-selection under projective measurements \cite{ADL} and in the theory of weak value measurements \cite {WeakMeasurement}, while the idea of a complete description of a quantum system at any time during a sequence of measurements \cite{Aharonov} has found a general dynamical formulation in the so-called past quantum state (PQS) \cite{KlausPRL,KlausPRA}. The PQS provides the probability distribution of the outcome of any general measurement on a quantum system at time $t$ conditioned on the preparation of the system and measurements on it both before and after $t$. The PQS has been demonstrated to yield better predictions than the usual conditional density matrix in trajectory simulations of the  photon number evolution in a cavity \cite{HarochePRA}, the excitation and emission dynamics of a superconducting qubit \cite{Murch}, and the motional state of a mechanical oscillator~\cite{Albert}.

Here, we show that for a metrologically relevant macroscopic atomic spin system, the standard quantum limit (SQL) determined by the atom projection noise can be surpassed, by conditioning the measurement result on  previous and later measurements on the system. The incorporation of later measurements supplement the well-established measurement-based entanglement generation protocol  ~\cite{Kitagawa,Kasevich,James,Appel,Vladan,Jessen,GH}, and more importantly, it improves on the quantum noise reduction.  We report here a conditional noise reduction of 5.6 dB, and spin squeezing of 4.5 dB using the Wineland criterion, for a collective spin of $1.87\times10^{11}$ hot atoms in a vapor cell, which corresponds to an angular spin variance of $4.6\times10^{-13} rad^2$, the best for a squeezed state up to date.
This work illustrates a new aspect of the quantum trajectory description with prospects for improvement of the already precise measurements with vapor cells for magnetometry \cite{Budker-book,Romalis,Romalis1,Polzik1}, fundamental symmetry test \cite{SSRMP,Romalis2,CPT} and gravitational wave detection \cite{Polzik2}.

\begin{figure}
\centering
\includegraphics[width=0.5\textwidth]{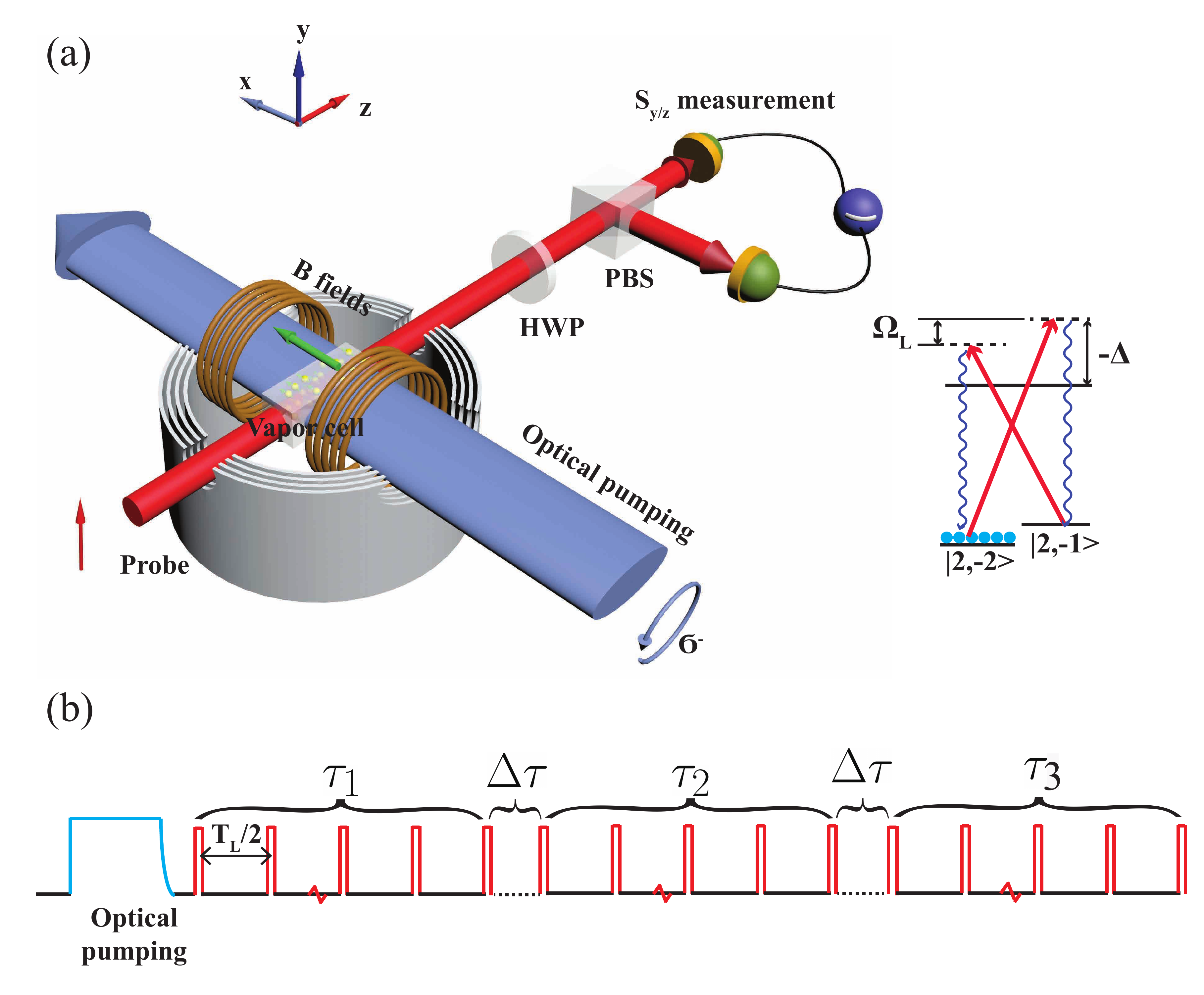}
\caption{\label{Fig:setup}\textbf{Experimental setup}.
\textbf{(a)} \textbf{Schematics}.  A paraffin coated $20\text{mm}\times7\text{mm}\times7\text{mm}$ rectangular vapor cell at $53^{\circ}$C resides inside a four-layer magnetic shielding to screen ambient magnetic field. The coherent spin state (CSS)  is created by optical pumping, with a pump laser tuned to the Rb D1 transition $5S_{1/2}F=2\rightarrow5P_{1/2}F'=2$ and a repump laser stabilized to the Rb D2 transition $5S_{1/2}F=1\rightarrow5P_{3/2}F'=2$, both with $\sigma ^{-}$ circular polarization along the $x$ direction. A magnetic field (along the \emph{x} direction) of $0.71~\text{G}$ is applied to induce a ground-state Zeeman splitting of $\Omega_L\simeq 2\pi\times500~\text{kHz}$ (i.e., Larmor frequency) and to hold the collective spin. A linearly polarized off-resonant D2 laser beam, propagating in the \emph{z} direction, probes the quantum fluctuations of the spin. The Stokes component $S_{y}$ is measured using a balanced polarimetry scheme and detected at the Larmor frequency $\Omega_L$  by a lock-in amplifier. \textbf{(b)} \textbf{Pulse sequence}. The pump lasers prepare the atoms in the CSS, and are then turned off adiabatically, followed by the stroboscopic probe pulses spaced by half the Larmor period $T_L$. The first part (pulse duration $\tau_1$) of the probe, called squeezing pulse, creates entanglement between $S_{y}$ and $J_{z}$. $J_{z}$ is squeezed through the detection of $S_{y}$, and the second part (pulse duration $\tau_2$), called the verification pulse, verifies the squeezing. The state is further probed (squeezed) for a duration $\tau_3$. The gap time $\Delta\tau=0.3$ ms between the three probe periods is to avoid interference from the lock-in amplifier.}
\end{figure}

Consider a collective atomic spin given by the sum of the total angular momenta of individual atoms, $\hat{J}_i=\sum_{k}\hat{j}_{i}^k$, with $i=x, y, z$.  The macroscopic spin orientation $J_x$ is along the applied bias magnetic field $B$, and the collective spin components $\hat{J}_{y,z}$ oscillate in the lab frame at the Larmor frequency $\Omega_L$. In the rotating frame, they obey the commutation relation $\left [ \hat{J}_{y},\hat{J}_{z} \right ]=iJ_{x} (\hbar=1)$. 

The QND measurement of the collective atomic spin is realized by coupling the atomic ensemble to a light beam with the off-resonant Faraday interaction depicted in Eq. (\ref{Hamiltonian}), such that a direct measurement on the transmitted field provides information about the atomic spin.
The Hamiltonian of the far off-resonance atom-light interaction in our experiment is \cite{RMP, GH}
\begin{equation}\label{Hamiltonian}
 \begin{aligned}
  \ \hat{H}_{int}=(\sqrt{2}\kappa/\sqrt{N_{ph}N_{at}})\hat{J}_{z}\hat{S}_{z},
 \end{aligned}
\end{equation}
where $N_{ph}$ is the number of photons in the pulse duration of $\tau$, and $N_{at}$ is the atom number. $\hat{S}_{z}$ is the Stokes operator of the probe light, relating to the photon number difference between $\sigma ^{+}$ and $\sigma ^{-}$ polarization. The coupling constant $\kappa^2\propto d_0\eta\propto N_{ph}N_{at}$ characterizes the measurement strength in quantum nondemolition (QND) detection, with $d_0$ the resonant optical depth and $\eta$ the atomic depumping rate causing decay of the collective spin.

We use an ensemble of $10^{11}$ $^{87}\text{Rb}$ atoms contained in a paraffin coated vapor cell \cite{Balabas} as shown in Figure~\ref{Fig:setup}. The coating provides a spin-protecting environment, enabling the high-performance optical pumping and promising the long spin coherence time to reduce the information loss due to decoherence. The atoms are initially prepared in the state $5S_{1/2}\left | F=2, m_F =-2\right \rangle$ (defined by the quantization axis \emph{x}) by optical pumping propagating along the \emph{x}-direction parallel to the \emph{B} field. We achieve up to $97.9\%$ polarization of the spins, resulting in a $6\%$ increase of the measured variance compared to the fully polarized Coherent Spin State (CSS). The quantum fluctuations of the spin is probed by a linearly-polarized off-resonant D2 laser beam, propagating in the \emph{z} direction. The projection noise limit is calibrated by measuring the noise of the collective spin of the unpolarized sample, which is 1.25 times that of CSS state. The QND measurement of the spin component $\hat{J}_z$ is achieved by implementing the stroboscopical quantum back-action evasion protocol~\cite{GH} (i.e. modulating the measurement intensity at twice the Larmor frequency with an optimal duty factor of $14\%$).

To describe the atomic system and its collective spin fluctuations during the optical probing, we apply the general quantum theory of measurements. To account for the quantum state conditioned on both prior and posterior probing of a quantum system, we consider a system subject to three subsequent measurement processes. Each measurement $(i)$ is described as a general positive-operator valued measurement (POVM) with a set of operators $\left \{ \hat{\Omega}^{(i)}_m \right \}$ associated with the measurement outcome $m$ and fulfilling $\sum_m\hat{\Omega}_m^{(i)\dagger}\hat{\Omega}^{(i)}_m=\hat{\mathbb{I}}$. For a system represented by the density matrix $\rho$ at the time of a measurement, the probability of measuring outcome $m$ is,
\begin{equation}\label{priorP}
Pr^{(i)}(m)=\text{Tr}(\hat{\Omega}^{(i)}_m\rho \hat{\Omega}_m^{(i)\dagger}),
\end{equation}
and the resulting conditional state reads,
\begin{equation}\label{condstate}
\rho|_m = \hat{\Omega}^{(i)}_m\rho\hat{\Omega}_m^{(i)\dagger}/Pr^{(i)}(m).
\end{equation}
Assuming no further dynamics between the measurements, we can evaluate the joint probability that three subsequent measurements, described by $\left \{ \hat{\Omega}^{(i)}_m \right \}$ yield outcomes $m_1$, $m_2$ and $m_3$, as
\begin{equation}\label{jointprob}
Pr(m_1,m_2,m_3) = \text{Tr}(\hat{\Omega}^{(3)}_{m_3}\hat{\Omega}^{(2)}_{m_2}\hat{\Omega}^{(1)}_{m_1} \rho \hat{\Omega}_{m_1}^{(1)\dagger}\hat{\Omega}_{m_2}^{(2)\dagger}\hat{\Omega}_{m_3}^{(3)\dagger}).
\end{equation}
Reading this equation from the inside and out, it can be factored into {\it (1)} the probability of obtaining the first outcome $m_1$, {\it (2)} the probability of obtaining outcome $m_2$ in the state conditioned on the first outcome, and {\it (3)} the probability of obtaining outcome $m_3$ in the state conditioned on the first two outcomes. This is equivalent with the conventional evolution of quantum trajectories, where the quantum state, and hence the probability of a measurement outcome depends on previous measurements. But, the joint probability distribution \eqref{jointprob} also permits evaluation of the probability of, say the second measurement, conditioned on the outcome of the first and the last one,
\begin{equation}\label{simplepast}
Pr(m_2|m_1,m_3) = Pr(m_1,m_2,m_3)/\sum_{m_2'} Pr(m_1,m_2',m_3),
\end{equation}
where $m_1,m_3$ are fixed to the observed values and the denominator is merely a normalization factor.

Using Eq. \eqref{jointprob} and the cyclic permutation property of the trace, we can write this probability  as  \cite{KlausPRL}
\begin{equation}\label{pastP}
Pr_p(m_2,t)=\frac{\text{Tr}(\hat{\Omega}^{(2)}_{m_2} \rho|_{m_1} \hat{\Omega}_{m_2}^{(2)\dagger} E|_{m_3})}{\sum_{m'}\text{Tr}(\hat{\Omega}^{(2)}_{m'}\rho|_{m_1}\hat{\Omega}_{m'}^{(2)\dagger}E|_{m_3})},
\end{equation}
where $\rho|_{m_1}$ is the state conditioned on the first measurement, cf. \eqref{condstate}, and   $E|_{m_3}=\hat{\Omega}_{m_3}^{(3)\dagger} \hat{\Omega}^{(3)}_{m_3}$.

We observe that the conventional expression for the outcome probabilities \eqref{priorP} conditioned only on the prior evolution of $\rho$ is supplemented with the operator $E$ which depends only on the later measurement outcomes. The same formalism applies to cases with continuous sequences of measurements occurring simultaneously with Hamiltonian and dissipative evolution. Examples of how the operators $\rho(t)$ and $E(t)$ evolve to time $t$ from the initial and final time, respectively, are given in Ref. \cite{KlausPRL}.

The specific form of the POVM operators and their action on the quantum states in our experiments can be derived explicitly in a simplified form because our system dynamics is restricted to Gaussian states. This follows from the Holstein-Primakoff transformation that maps the spin operators perpendicular to the large mean spin on the canonical position and momentum operators, $\hat{x}_A=\hat{J}_{y}/\sqrt{\left | \left \langle J_x \right \rangle \right |}$ and $\hat{p}_A=\hat{J}_{z}/\sqrt{\left | \left \langle J_x \right \rangle \right |}$. The CSS with all atoms in $\left |F,m_F=-F\right \rangle$, characterized by $\text{Var}(\hat{J}_{y})=\text{Var}(\hat{J}_{z})=J_x/2=N_{at}F/2$, is equivalent to the Gaussian ground state of a harmonic oscillator, and an excitation with the ladder operator $\hat{b}^{\dagger}$  corresponds to a quantum of excitation, also called a polariton, distributed symmetrically among all the atoms \cite{GH}. Similar canonical operators, $\hat{x}_L=\hat{S}_{y}/\sqrt{\left | \left \langle S_x \right \rangle \right |}$ and $\hat{p}_L=\hat{S}_{z}/\sqrt{\left | \left \langle S_x \right \rangle \right |}$ and Gaussian states describe the probe field degrees of freedom.

The measurement operator $\hat{\Omega}_m$ (\ref{priorP}) acting on the atomic state upon readout of the value $m$ of the field quadrature $\hat{x}_L$ is given by,  $\hat{\Omega}_{m}=\int \psi_{\hat{x}_L}(m-\kappa a)| a \rangle\langle a |_{\hat{p}_A}da$, where
$\psi_{\hat{x}_L}(m)=\frac{1}{\pi^{1/4}}\exp(-\frac{m^2}{2})$ characterizes the quadrature distribution of the input coherent state of the probe laser beam.

For two successive QND measurements with coupling strengths $\kappa_1$ and $\kappa_2$, the POVM formalism shows that the second outcome is governed by a Gaussian distribution with a mean value conditioned on the first outcome,
\begin{equation}\label{dis_nopast}
\begin{aligned}
Pr(m_2 |m_{1})=\frac{1}{\sqrt{\pi}\sigma} \exp\left(-\frac{\left[m_2-\frac{\kappa_2m_1\kappa_1}{1+\kappa_1^2}\right]^2}
{2\sigma^2}\right).
\end{aligned}
\end{equation}
Here the variance $\sigma^2=\frac{1}{2}+\frac{1}{2}\frac{\kappa_2^2}{1+\kappa_1^2}$ is composed of a contribution $\frac{1}{2}$ from the light shot noise, and a contribution from the atomic spin, which is reduced by the conditional spin squeezing by the first measurement with strength $\kappa_1$.

If the spin oscillator is further probed by a third QND pulse with coupling strength $\kappa_3$ and measurement outcome $m_3$, the conditional probability for the outcome of the middle measurement is obtained as
\begin{equation}\label{dis_past}
\begin{aligned}
Pr(m_2 |m_1,m_3)=\frac{1}{\sqrt{\pi}\sigma_p}\exp\left(-\frac{\left[m_2-\frac{\kappa_2(m_1\kappa_1+m_3\kappa_3)}{1+\kappa_1^2+\kappa_3^2}\right]^2}
{2\sigma_p^2}\right).
\end{aligned}
\end{equation}
The past probability yields a Gaussian distribution with variance $\sigma_p=\frac{1}{2}+\frac{1}{2}\frac{\kappa_2^2}{1+\kappa_1^2+\kappa_3^2}$. The reduction by $1+\kappa_1^2+\kappa_3^2$ shows that the incorporation of the information from later measurements has a similar effect as if we had increased the coupling strength of the first probing from $\kappa_1^2$ to $\kappa_1^2+\kappa_3^2$.

\begin{figure}
\centering
\includegraphics[width=8cm]{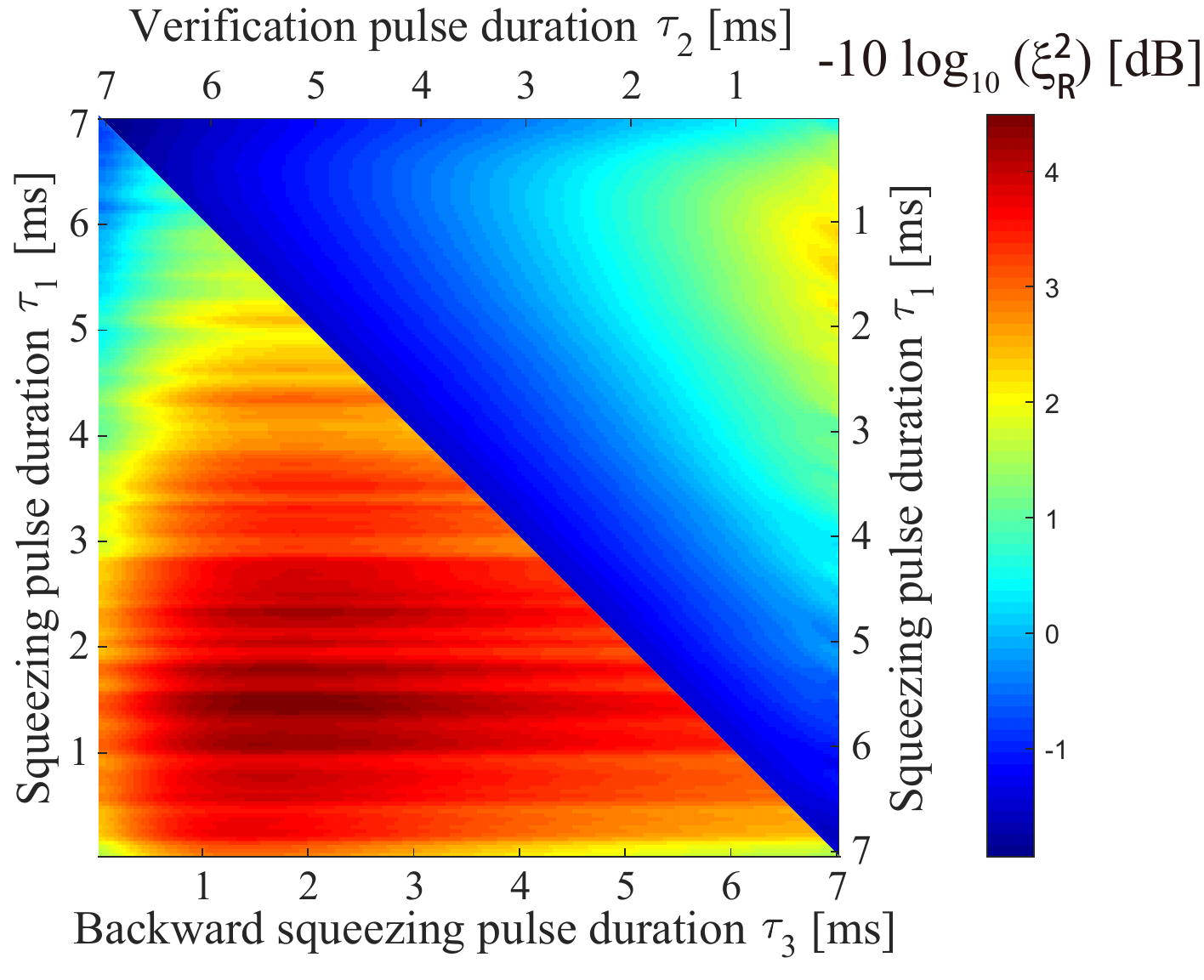}
\caption{\textbf{Experiment results.} Lower diagonal: degree of spin squeezing (denoted by color) of the three-pulse scheme, for various time durations of the $1^{st}$ and $3^{rd}$ pulses. The duration of the $2^{nd}$ probe pulse is 0.037 ms. Overall, the squeezing is better for shorter verification pulse $\tau_2$, and reaches its maximum value of 4.5 dB at $\tau_1=1.4$ ms and $\tau_3=1.7$ ms, balancing the increased dispersive interaction with the high photon number and the loss of polarization due to spontaneous emission. Upper diagonal: detected spin squeezing by the traditional squeezing and verification two-pulse scheme as function of $\tau_1$ and $\tau_2$. The best squeezing is 2.3 dB. The probe laser has an average power of 1.18 mW in both experiments. $\xi_{R} ^{2}$ is the squeezing parameter by the Wineland criterion.}
\label{squeezing}
\end{figure}

Experimentally, for the normal two-pulse scheme of forward conditioning QND, we achieve a minimum spin squeezing of $2.3\pm 0.2 ~\text{dB}$ (Figure \ref{squeezing} upper diagonal) by the Wineland criterion \cite{Wineland} for $\tau_1=1.23$ ms, and a conditional noise reduction of about 4.3 dB, in good agreement with the theoretical prediction. As predicted by \eqref{dis_past}, in stark contrast, for the three-pulse scheme which extracts the full information from the full measurement record by the past quantum state, we observe an improved conditional noise reduction of about 5.6 dB, and spin squeezing of $4.5\pm0.40~\text{dB}$ (Fig.~\ref{squeezing} lower diagonal) by the Wineland criterion, for $\tau_1=1.4$ ms and $\tau_3=1.7$ ms. As the dissipation induced by all pulses degrade the squeezing, we measure the highest squeezing for the shortest possible verification pulse $\tau_2=0.037$ ms.

The main reason that the probing before and after the verification pulse sequence yields stronger squeezing than an initial longer probing sequence, is the decoherence of the spins. Firstly, due to decoherence, the spin squeezing is gradually lost, and measurement results obtained during the early stages of the squeezing ($1^{st}$) pulse sequence will be less correlated with the spin ensemble at the time of the verification ($2^{nd}$) pulse. If we instead postpone these measurements to occur in the $3^{rd}$ pulse sequence immediately after the verification pulse, the correlations will be stronger, i.e., the conditional variance will be lower. Secondly, the large average spin component $J_x$ is reduced during probing, weakening the squeezing according to the Wineland criterion. With the retrodicted squeezing, the spin variance is measured relative to the mean spin at the time of the verification pulse, which has not yet suffered the reduction due to the $3^{rd}$ pulse sequence.

\begin{figure}
\centering
\includegraphics[width=0.45\textwidth]{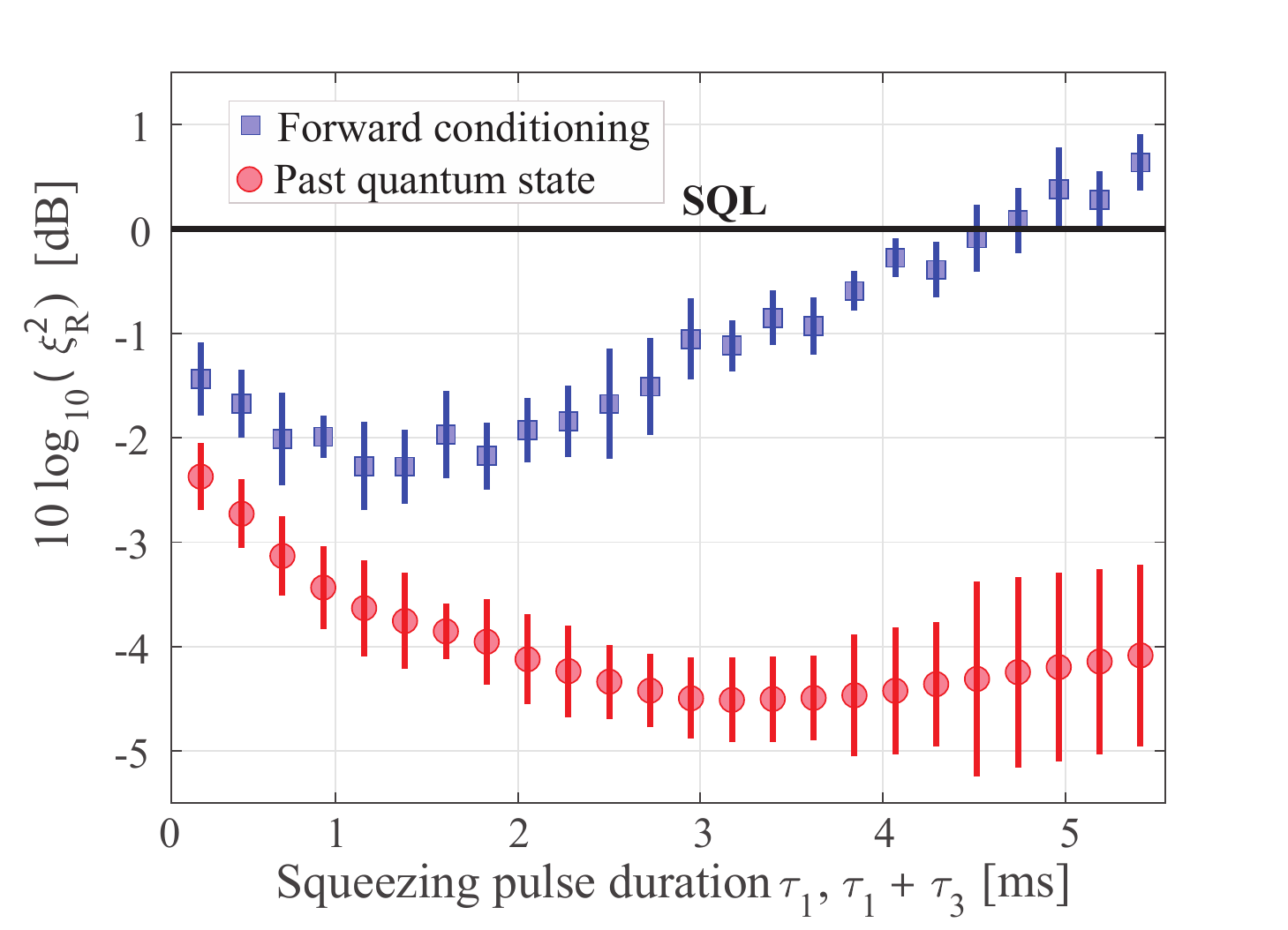}
\caption{\label{compare2_3pulses} \textbf{Squeezing vs. the total squeezing pulse duration for comparison of two- and three-pulse schemes.} The horizontal axis shows $\tau_1$ for the two-pulse scheme (forward conditioning) and $\tau_1+\tau_3$ for the three-pulse scheme (past quantum state protocol). $\tau_2$ is 0.037 ms for both curves. The attainable squeezing for the three-pulse scheme has a lower minimum and better long time behavior than those for the two-pulse scheme. The error bar is derived from 10 identical experiments, with each consisting of 10000 repetitions of the pulse sequence shown in Fig.1(b).}
\end{figure}

As shown in Fig.\ref{compare2_3pulses}, even if we hold the total duration of squeezing equal for both schemes, the attainable squeezing using both the information before and after the $2^{nd}$ pulse is better than that of using only the information before the $2^{nd}$ pulse. This improved squeezing result can be used to improve the measurement precision of a temporary shift of $J_z$, occurring , e.g., due to a local time-varying RF magnetic field (at frequency $\Omega_L$) around the time of the $2^{nd}$ pulse.

We have demonstrated quantum noise reduction below the SQL for a macroscopic collective spin of $10^{11}$ atoms, containing 1000 times more atoms than previous squeezed states. This work thus sets a new record on how large (in terms of the number of spins) a physical system can be while still benefiting from quantum measurements. We theoretically and experimentally proved that the past quantum state approach allows for a sharper prediction for the projective measurement on a spin oscillator, than the usual one via forward conditioning. Atoms constitute ideal probes for a number of physical phenomena with high sensitivity\cite{SSRMP,Romalis2,CPT}, and our retrodiction procedure may significantly impact their practical applications as quantum sensor, such as magnetometers \cite{Budker-book,Romalis,Romalis1,Polzik1}. In particular, the retrodicted evolution of physical systems may offer insight and allow precision estimation of time dependent perturbations \cite{Tsang}, applicable, e.g., to force sensing with mechanical oscillators \cite{Polzik2,OptoRMP}.

\section*{\textbf{Acknowlegements}}
This work is supported by National Key Research Program of China under Grant No. 2016YFA0302000 and No. 2017YFA0304204, and NNSFC under Grant No. 61675047 and No. 91636107. K. M. acknowledges support from the Villum Foundation. H. S. acknowledges the financial support from the Royal Society Newton International Fellowship (NF170876) of UK.
\end{document}